\documentclass[prc,amsmath,amssymb,showpacs,twocolumn]{revtex4}
\usepackage{epsfig,color}

\begin{document} 

\title{Low-energy behavior of $E2$ strength functions}

\author{R. Schwengner} 
\affiliation{Helmholtz-Zentrum Dresden-Rossendorf, 01328 Dresden, Germany}

\date{\today} 

\begin{abstract}
Electric quadrupole strength functions have been deduced from averages of a
large number of $E2$ transition strengths calculated within the shell
model for the nuclides $^{94}$Mo and $^{95}$Mo. These strength functions are
at variance with phenomenological approximations as provided by the Reference
Input Parameter Library RIPL-3 for calculations of reaction rates on the basis
of the statistical model.
\end{abstract}


\maketitle

\section{Introduction}
\label{sec:intro}

Photonuclear reactions and the inverse radiative-capture reactions, in
particular radiative neutron capture, play a central role in the synthesis of
heavy elements in various stellar environments \cite{Arnould07,Kaeppler11} and
also in next-generation nuclear technologies, such as the transmutation of
long-lived nuclear waste \cite{Arnould07,Chadwick11}. As these reactions
include the excitation and deexcitation of nuclear states at high excitation
energy and large level density, the so-called quasicontinuum of states,
the statistical reaction theory is the basis for calculations of rates
of these reactions. A critical input to such calculations is photon strength
functions that describe average electromagnetic transition strengths.
Modifications of the strength functions can change reaction rates considerably.
For example, modifications of the electric dipole strength function have
drastic consequences for the abundances of elements produced via neutron
capture in the r-process occurring in violent stellar events \cite{gor98}.

In the calculations using statistical codes (e.g. TALYS \cite{kon05}), usually
electric dipole ($E1$), magnetic dipole ($M1$), and electric quadrupole ($E2$)
strength functions are taken into account. In the energy range below the
particle-separation energies, which is relevant for radiative capture
reactions, the dipole strength function is dominated by the low-energy tail of
the isovector electric giant dipole resonance (GDR). The GDR is considered as a
collective vibration of the neutron system against the proton system. The
damping of the vibration is described by a Lorentz curve as a function of the
photon energy \cite{bri55,axe62,cap10}. Combinations of two or three Lorentz
curves are used to describe the double or triple humps of the GDR caused by
quadrupole and triaxial deformation of the nuclei \cite{boh75,eis75,jun08}.
Such a parametrization gives a good description of the experimental
photoabsorption cross section
$\sigma_\gamma$ = 3 $(\pi \hbar c)^2$ $E_\gamma$ $f_1(E_\gamma)$ of nuclei in
the ground state. The so-called Brink-Axel hypothesis \cite{bri55,axe62}
expresses the assumption that the strength function does not depend on the
excitation energy. This means that the strength function describing the
absorption of photons is identical with the one for the emission of photons
from highly excited states, for example following neutron capture. The
Generalized Lorentzian (GLO) \cite{kop90} includes a correction to the
Standard Lorentzian (SLO) \cite{bri55,axe62}, which accounts for the
temperature of the nucleus emitting the photons.
The magnetic dipole $(M1)$ contribution to the strength function used in
statistical-reaction codes is also approximated by a Lorentz curve with
parameters derived from systematics \cite{cap10}. This curve accounts for the
spin-flip mode that appears around 8 MeV \cite{hey10}.

In several experiments, deviations from the phenomenological strength functions
have been observed. A bump of the $M1$ strength around 3 MeV in
deformed nuclei is generated by the scissors mode, which is interpreted as a
small-amplitude rotation of the neutron system against the proton system
\cite{hey10}. After it had been well established  in the absorption spectra of
the ground state, it was recently also identified in the emission from highly
excited states \cite{gut12}.

An enhancement of $E1$ strength has been found in the energy region from about
6 MeV up to the respective neutron-separation. This additional strength on top
of the low-energy tail of the GDR is considered as the pygmy dipole resonance
(PDR) which is interpreted as the vibration of excessive neutrons against the
symmetric $N = Z$ neutron-proton system. A review of experimental studies of
the PDR can be found in Ref.~\cite{sav13}.

In contrast to the Lorentz curves used for the $E1$ and $M1$ strength
functions, which decrease toward $E_\gamma$ = 0, an increase of the dipole
strength below 3 MeV toward low $\gamma$-ray energy has been found in several
nuclides in the mass range from $A \approx$ 50 to 100, such as $^{56,57}$Fe
\cite{voi04}, $^{60}$Ni \cite{voi10}, various Mo isotopes \cite{gut05}, and
$^{105,106}$Cd \cite{lar13}. Neither of these measurements were able to
distinguish clearly between $E1$ and $M1$ strength, although an indication for
an $M1$ character of the low-energy enhancement was discussed for the case of
$^{60}$Ni \cite{voi10}. In an analysis of $M1$ strength functions deduced from
shell-model calculations of a large number of transitions in the isotopes 
$^{90}$Zr, $^{94}$Mo, $^{95}$Mo, and $^{96}$M \cite{sch13} we showed that
the low-energy enhancement of the dipole strength can be explained by $M1$
transitions between many close-lying states of all considered spins located
above the yrast line in the transitional region to the quasi-continuum of
nuclear states. Inspecting the wave functions, one finds large $B(M1)$ values
for transitions between states containing a large component of the same
configuration with broken pairs of both protons and neutrons in high-$j$
orbits. The large $M1$ matrix elements connect configurations with the spins of
high-$j$ protons re-coupled with respect to those of high-$j$ neutrons to the
total spin $J_f = J_i, J_i \pm 1$.

In an alternative work the low-energy enhancement could be described by $E1$
strength generated by the thermal coupling of quasiparticles to the continuum
of unbound states \cite{lit13}. This effect appears at temperatures above 1.4
MeV, whereas experimentally deduced values and values predicted by the
constant-temperature and Fermi-gas models are below 1.0 MeV \cite{gut05,egi09}.

For the $E2$ strength function a Lorentz curve is recommended as well
in the RIPL-3 reaction data base \cite{bel06} with the following parameters:
energy of the maximum $E_{\rm max} = 63 A^{-1/3}$ MeV, width
$\Gamma = 6.11 - 0.021 A$ MeV, and maximum of the corresponding cross section
$\sigma_{\gamma, \rm max} = 0.00014 Z^2 E_{\rm max} A^{-1/3} \Gamma^{-1}$ mb.
The Lorentz function in combination with the factor $E_\gamma^{-2L+1}$ produces
an unrealistic pole at $E_\gamma$ = 0. An experimental test of the real
behavior of the $E2$ strength function at low transition energy has not been
feasible so far. However, model calculations may gain information about the
$E2$ strength function at low energy. As the shell-model calculations just
mentiond were successful in describing the low-energy enhancement of the $M1$
strength observed in various experiments \cite{sch13}, these calculations are
expected to predict also the low-energy behavior of the $E2$ strength functions
in the considered nuclei near $N$ = 50.

\section{Shell-model calculations}
\label{sec:shell}

The present work presents shell-model calculations of $E2$ transition
strengths in $^{94}$Mo and $^{95}$Mo. The calculations were performed by means
of the code RITSSCHIL \cite{zwa85} using a model space composed of the
$\pi(0f_{5/2}, 1p_{3/2}, 1p_{1/2}, 0g_{9/2})$ proton orbits and the
$\nu(1p_{1/2}, 0g_{9/2}, 1d_{5/2})$ neutron orbits relative to a $^{66}$Ni
core. The configuration space was tested in detail in earlier shell-model
studies of nuclei with $N = 46 - 54$ \cite{sch09,sch022,sch95,
sch982,sch06,win93,win94,rei95,ste00,zha04,jun98,jun99,ste01,rai02,ste02} and
was found appropriate for the description of level energies as well as
$M1$ and $E2$ transition strengths in nuclides around $A$ = 90. 
As a further test, a comparison of the energies of yrast and yrare levels 
in $^{94}$Mo and $^{95}$Mo from the present calculation with the
experimental ones shows an agreement within 300 keV. 

The calculations included states with spins from $J$ = 0 to 10 for $^{94}$Mo
and from $J$ = 1/2 to 21/2 for $^{95}$Mo. Two protons were allowed to be lifted
from the $1p_{3/2}$, $1p_{1/2}$ orbits to the $0g_{9/2}$ orbit and two neutrons
from the $0g_{9/2}$ orbit to the $1d_{5/2}$ orbit. This resulted in
configuration spaces with dimensions of up to about 16000. For each spin the
lowest 40 states were calculated. The reduced transition probabilities $B(E2)$
were calculated for all transitions from initial to final states with energies
$E_f < E_i$ and spins $J_f = J_i, J_i \pm 1, J_i \pm 2$. For the minimum and
maximum $J_i$, the cases $J_f < J_i$ and $J_f > J_i$, respectively, were
excluded. This resulted in more than 36000 $E2$ transitions for each parity
$\pi = +$ and $\pi = -$, which were sorted into 100 keV bins according to the
excitation energy of the initial state $E_i$ or the transition energy
$E_\gamma = E_i - E_f$. The average $B(E2)$ value for one energy bin was
obtained as the sum of all $B(E2)$ values divided by the number of transitions
within this bin. Effective charges of $e_\pi = 1.5 e$ and $e_\nu = 0.5 e$ were
applied.

\section{Results}
\label{sec:results}

\begin{figure}
\epsfig{file=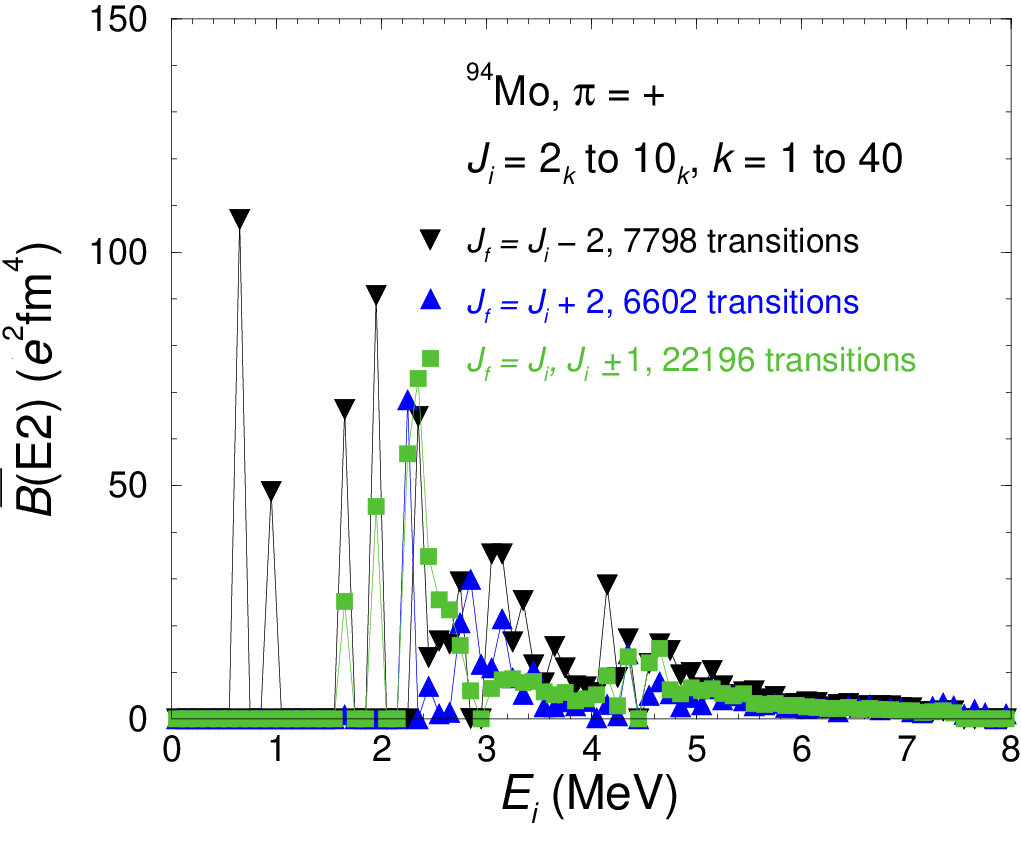,width=8cm}
\caption{\label{fig:94MoPExE2}(Color online) Average $B(E2)$ values in 100 keV
bins of excitation energy calculated for positive-parity states in $^{94}$Mo.}
\end{figure}

\begin{figure}
\epsfig{file=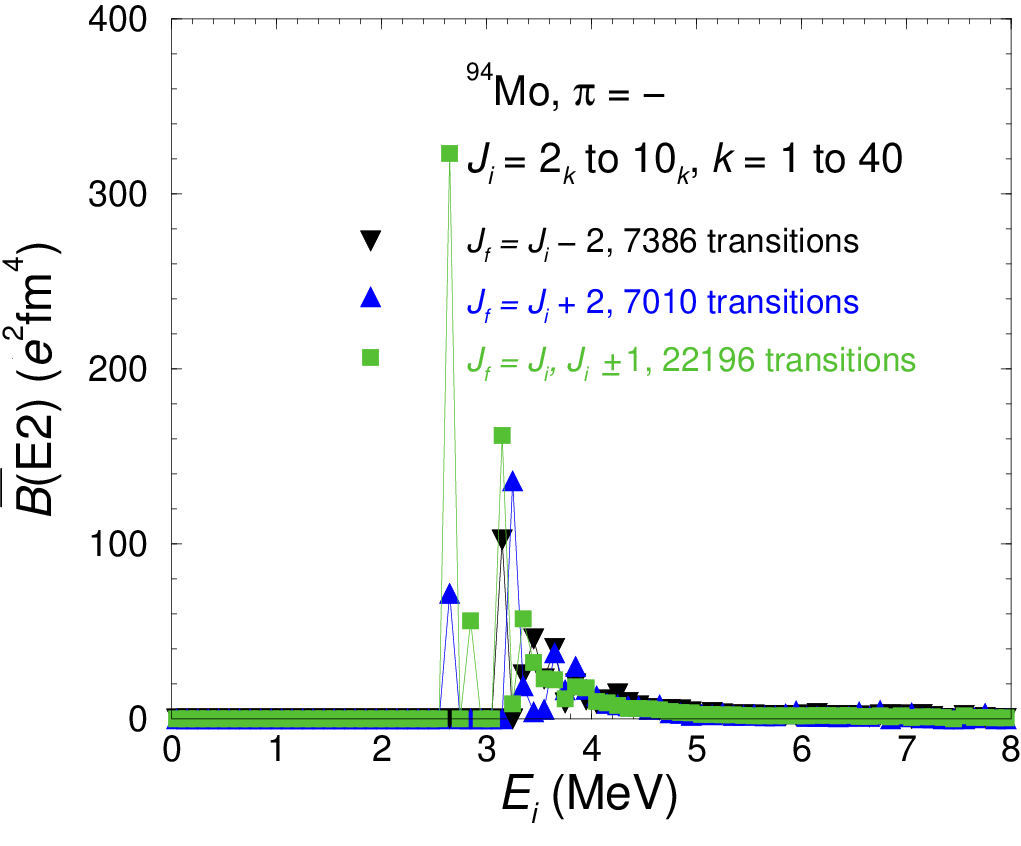,width=8cm}
\caption{\label{fig:94MoNExE2}(Color online) As Fig.~\ref{fig:94MoPExE2}, but
for negative-parity states.} 
\end{figure}

Average calculated $B(E2)$ values in 100 keV wide energy bins of initial 
excitation energy of positive-parity and negative-parity states in $^{94}$Mo
are shown in Figs.~\ref{fig:94MoPExE2} and \ref{fig:94MoNExE2}, respectively.
The $B(E2)$ values are separately shown for transitions with $J_f = J_i - 2$,
$J_f = J_i + 2$, and $J_f = J_i, J_i \pm 1$.


\begin{figure}
\epsfig{file=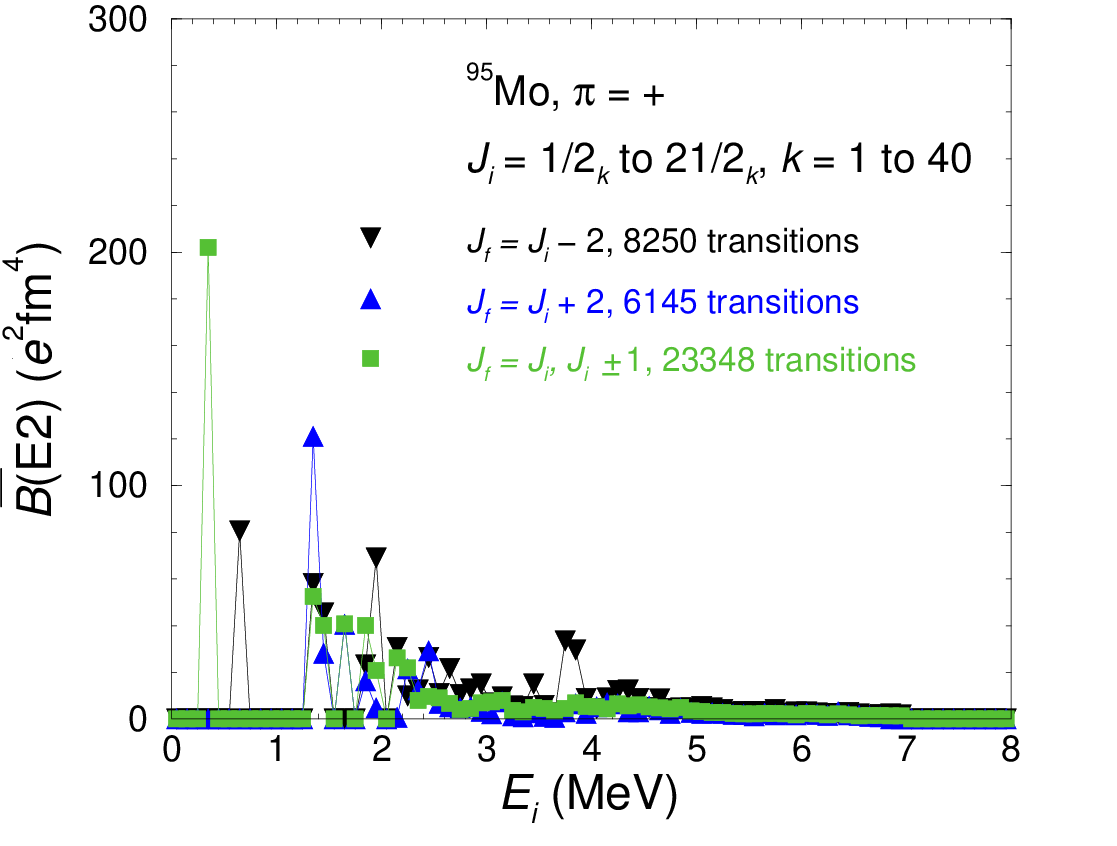,width=8cm}
\caption{\label{fig:95MoPExE2}(Color online) Average $B(E2)$ values in 100 keV
bins of excitation energy calculated for positive-parity states in $^{95}$Mo.}
\end{figure}
\begin{figure}
\epsfig{file=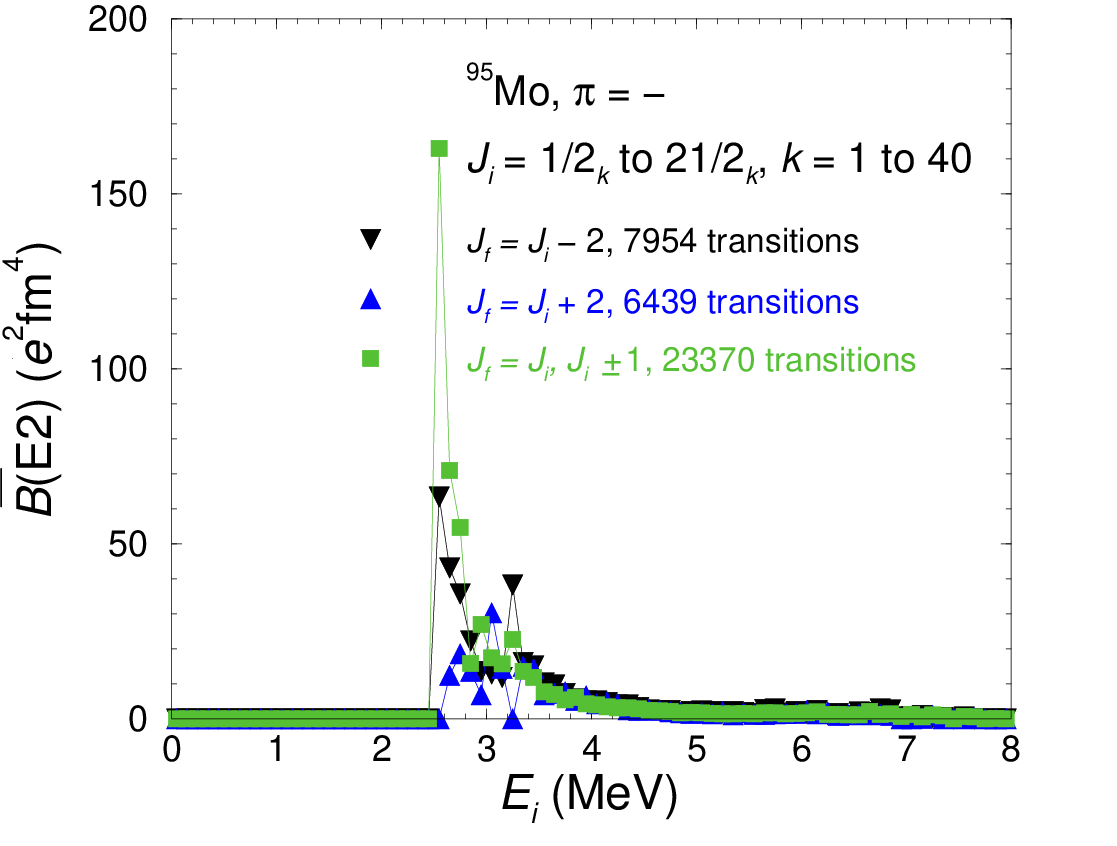,width=8cm}
\caption{\label{fig:95MoNExE2}(Color online) As Fig.~\ref{fig:94MoPExE2}, but
for negative-parity states.} 
\end{figure}

In Fig.~\ref{fig:94MoPExE2}, the peak at 0.65 MeV (energy bin from 0.6 to 0.7
MeV) arises from the $2^+_1 \rightarrow 0^+_1$ transition. The calculated
transition strength of $B(E2)$ = 107 $e^2$fm$^4$ is considerably smaller than
the experimental value of $B(E2)$ = 406(9) $e^2$fm$^4$ \cite{abr06}. This holds
also for the the $4^+_1 \rightarrow 2^+_1$ transition that dominates the peak
at 0.95 MeV. Its calculated value of $B(E2)$ = 49 $e^2$fm$^4$ has to be
compared with the experimental value of $B(E2)$ = 660(102) $e^2$fm$^4$ deduced
from Coulomb excitation \cite{abr06}. This comparison shows that collective
contributions to the lowest-lying (yrast) states are not fully accounted for in
the present configuration space. On the other hand, the calculated value of
$B(E2,4^+_2 \rightarrow 2^+_1)$ = 147 $e^2$fm$^4$, that predominates in the
peak at 1.95 MeV, fits exactly the experimental value of 
$B(E2,4^+_2 \rightarrow 2^+_1)$ = 147(23) $e^2$fm$^4$. This transition connects
a non-yrast $4^+$ state containing the main configuration
$\pi(0g_{9/2}^2) \nu(1d_{5/2}^2)$ with the $2^+$ yrast state including mainly
the configuration $\nu(1d_{5/2}^2)$. The configuration
$\pi(0g_{9/2}^2) \nu(1d_{5/2}^2)$ of two active proton and two
active neutron high-$j$ orbits is found to be the dominating configuration in
the many close-lying states above the yrast line which are connected by $M1$
transitions of large strengths \cite{sch13}. The $E2$ transitions between
states with $J_f = J_i, J_i \pm 1$ represent admixtures to those $M1$
transitions. As the states with the predominating four-particle configuration
contain little collectivity, the magnitudes of the calculated $B(E2)$ values
between them are considered more realistic than those of the stretched $E2$
transitions between low-spin yrast states, which is demonstrated by the
$B(E2,4^+_2 \rightarrow 2^+_1)$ value just mentioned. The lowest states
linked by a strong $M1$ transition are the $2^+_3$ and $2^+_1$ states. The
calculated $B(E2,2^+_3 \rightarrow 2^+_1)$ value of 87 $e^2$fm$^4$ compares
with an experimental value of 126(76) $e^2$fm$^4$ \cite{abr06}.

The distributions at higher $E_i$ from about 2.5 to 5.5 MeV include
contributions from transitions between many states with various spins. The
$B(E2)$ values generally decrease with increasing excitation energy, which is
also found for the experimental values compiled in Ref.~\cite{abr06}. In
Fig.~\ref{fig:94MoNExE2}, the peaks at 2.65 and 3.15 MeV are dominated by the
$E2$ admixtures to the $5^-_1 \rightarrow 4^-_1$ and $5^-_2 \rightarrow 4^-_2$
transitions. The $B(E2)$ distributions for the negative-parity states start at
higher excitation energy and decrease faster toward higher energy in comparison
with the ones for positive parity.

The values calculated for $^{95}$Mo, shown in Figs.~\ref{fig:95MoPExE2} for 
positive-parity states and in Fig.~\ref{fig:95MoNExE2} for negative-parity 
states, display a similar behavior. In Fig.~\ref{fig:95MoPExE2}, the peak at
0.35 MeV is caused by the $E2$ admixture to the $3/2^+_1 \rightarrow 5/2^+_1$
transition. The calculated value of $B(E2)$ = 202 $e^2$fm$^4$ compares with
the experimental value of $B(E2)$ = 554(28) $e^2$fm$^4$ \cite{bas10}.
The peak at 0.65 MeV corresponds to the value of
$B(E2,9/2^+_1 \rightarrow 5/2^+_1)$ = 81 $e^2$fm$^4$ compared with an
experimental value of $B(E2,9/2^+_1 \rightarrow 5/2^+_1)$ = 291(15) $e^2$fm$^4$
\cite{bas10}. Again, the calculated $B(E2)$ values of transitions between yrast
states underestimate the experimental values. In Fig.~\ref{fig:95MoNExE2}, the
peak formed by the values at 2.55, 2.65, and 2.75 MeV in the distribution of
values with $J_f = J_i, J_i \pm 1$ is caused by the $B(E2)$ strengths of the
$3/2^-_1 \rightarrow 5/2^-_1$, $7/2^-_1 \rightarrow 5/2^-_1$,
$7/2^-_2 \rightarrow 5/2^-_1$, and $5/2^-_2 \rightarrow 3/2^-_1$ transitions.
Main contributions to the peak formed at the same energies in the distribution
of $J_f = J_i - 2$ transitions arise from several $7/2^- \rightarrow 3/2^-$,
$9/2^- \rightarrow 5/2^-$, $11/2^- \rightarrow 7/2^-$, 
$13/2^- \rightarrow 9/2^-$, and $15/2^- \rightarrow 11/2^-$ transitions.

\begin{figure}
\epsfig{file=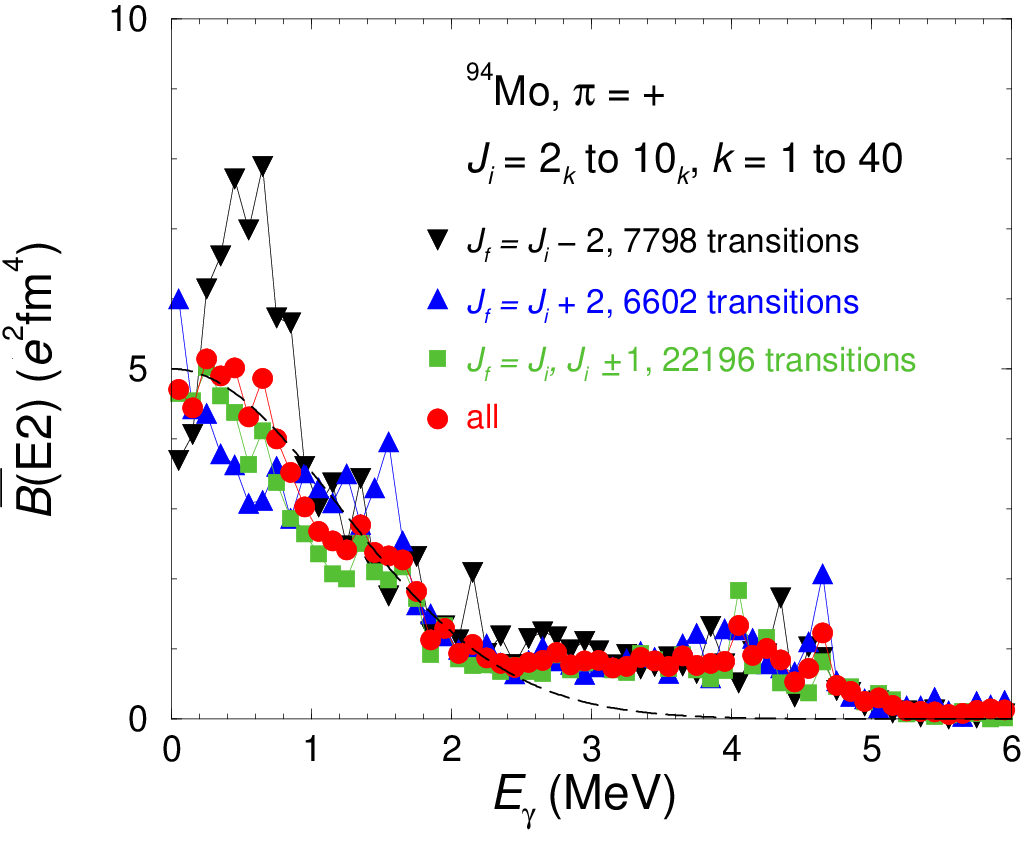,width=8cm}
\caption{\label{fig:94MoPEgE2}(Color online) Average $B(E2)$ values in 100 keV
bins of transition energy calculated for positive-parity states in $^{94}$Mo.
The black dashed curve is a Gau{\ss} curve with parameters given in the text.}
\end{figure}

\begin{figure}
\epsfig{file=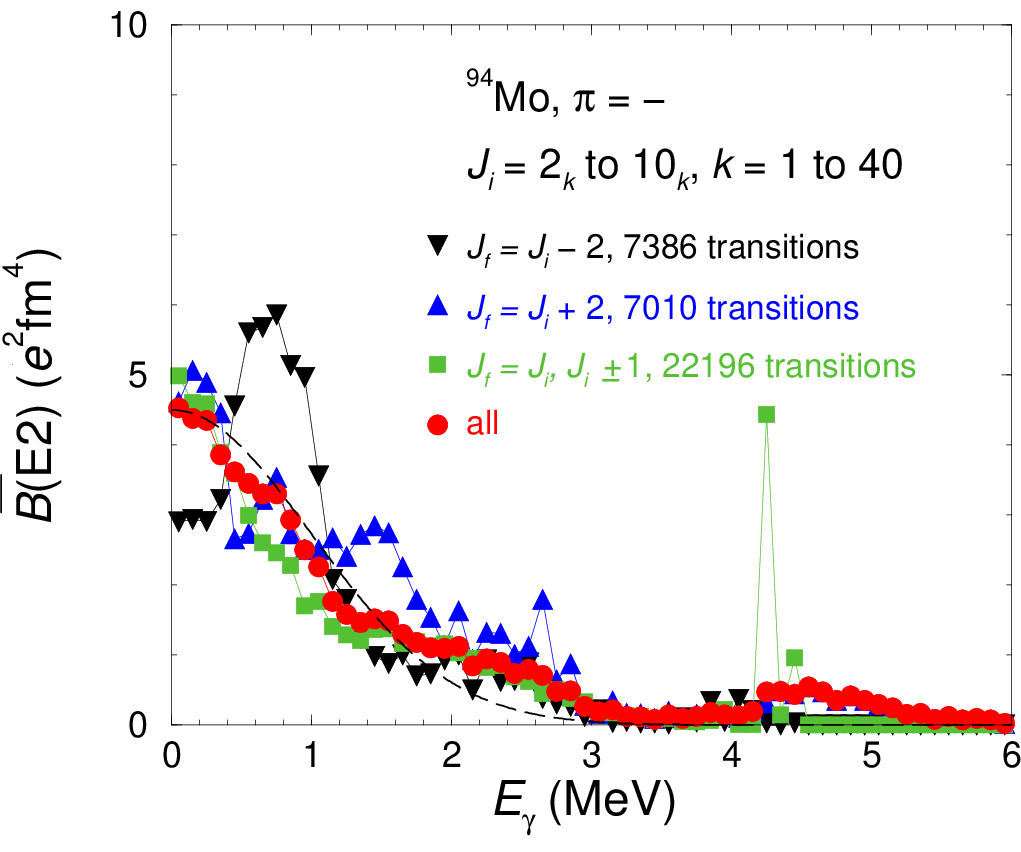,width=8cm}
\caption{\label{fig:94MoNEgE2}(Color online) As Fig.~\ref{fig:94MoPEgE2}, but
for negative-parity states.} 
\end{figure}

With regard to strength functions it is interesting to consider average $B(E2)$
values as a function of transition energy. Here, I will focus on the
low-energy part of the distributions of $\overline{B}(E2)$ values. As just
discussed, the calculated $B(E2)$ values of the low-energy transitions
connecting many close-lying states with the dominating four-particle
configuration $\pi(0g_{9/2}^2) \nu(1d_{5/2}^2)$ are roughly reproduced in
their magnitude and in particular the shape of the distributions at low
transition energy below about 2 MeV is therefore considered realistic.

The $\overline{B}(E2)$ values are shown in Fig.~\ref{fig:94MoPEgE2} for
positive-parity states and in Fig.~\ref{fig:94MoNEgE2} for negative-parity
states in $^{94}$Mo. For both parities, the $\overline{B}(E2)$ values of
stretched transitions with $J_f = J_i - 2$ peak in the energy region between
0.4 and 1 MeV, whereas the $\overline{B}(E2)$ values of the
$J_f = J_i + 2$ and $J_f = J_i, J_i \pm 1$ increase with a slope getting gentle
toward $E_\gamma$ = 0. The decrease toward high energy is followed by peaks
around 4.5 MeV for each parity. 
For positive-parity states shown in Fig.~\ref{fig:94MoPEgE2}, the peak in
the distribution of $J_f = J_i - 2$ transitions arises from transitions
depopulating high-lying $2^+$ to the $0^+_1$ and $0^+_2$ states, high-lying
$4^+$ to the $2^+_1$ and $2^+_2$ states, and so on. The peak seen for
$J_f = J_i + 2$ transitions is caused by transitions from high-lying
$0^+$ to the $2^+_1$ and $2^+_2$ states. The $\overline{B}(E2)$ values for
$J_f = J_i, J_i \pm 1$ transitions in the energy range between about 4 and 5
MeV belong to transitions from high-lying $1^+$ and $2^+$ states to the $2^+_1$
and $2^+_2$ states.
For negative-parity states shown in Fig.~\ref{fig:94MoNEgE2}, the peak in the
distribution of $J_f = J_i + 2$ states around 4.3 MeV is formed by transitions
from high-lying $0^-$ to the $2^-_1$ and $2^-_2$ states. 

Also shown in Figs.~\ref{fig:94MoPEgE2} and \ref{fig:94MoNEgE2} are the
distributions including all transitions of positive and negative parity,
respectively. These distributions are dominated by the behavior of the 
$\overline{B}(E2)$ values of the $J_f = J_i, J_i \pm 1$ transitions because of
their large number. The bump in the distribution of the $\overline{B}(E2)$
values of the $J_f = J_i - 2$ transitions is averaged out, which again shows
that in particular stretched $E2$ transitions between slightly collective yrast
states have a minor influence on the low-energy behavior of the $E2$ strength
functions. At energies below about 2 MeV, the $\overline{B}(E2)$ distributions
may be approximated by Gau{\ss} curves 
$\overline{B}(E2) = B_0 \exp(-E^2_\gamma/2\sigma^2)$ with 
$B_0$ = 5.0 $e^2$fm$^4$, $\sigma$ = 1.2 MeV for positive parity and 
$B_0$ = 4.5 $e^2$fm$^4$, $\sigma$ = 1.0 MeV for negative parity.

\begin{figure}
\epsfig{file=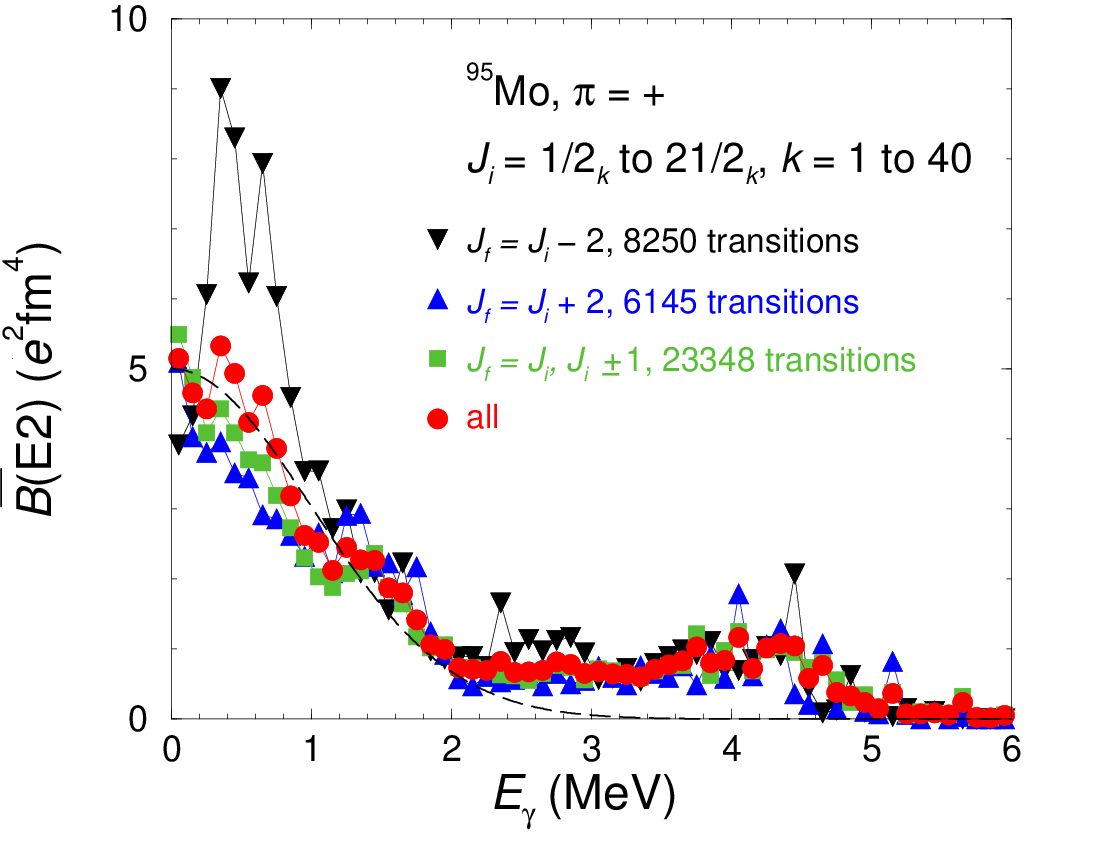,width=8cm}
\caption{\label{fig:95MoPEgE2}(Color online) Average $B(E2)$ values in 100 keV
bins of transition energy calculated for positive-parity states in $^{95}$Mo.
The black dashed curve is a Gau{\ss} curve with parameters given in the text.}
\end{figure}
\begin{figure}
\epsfig{file=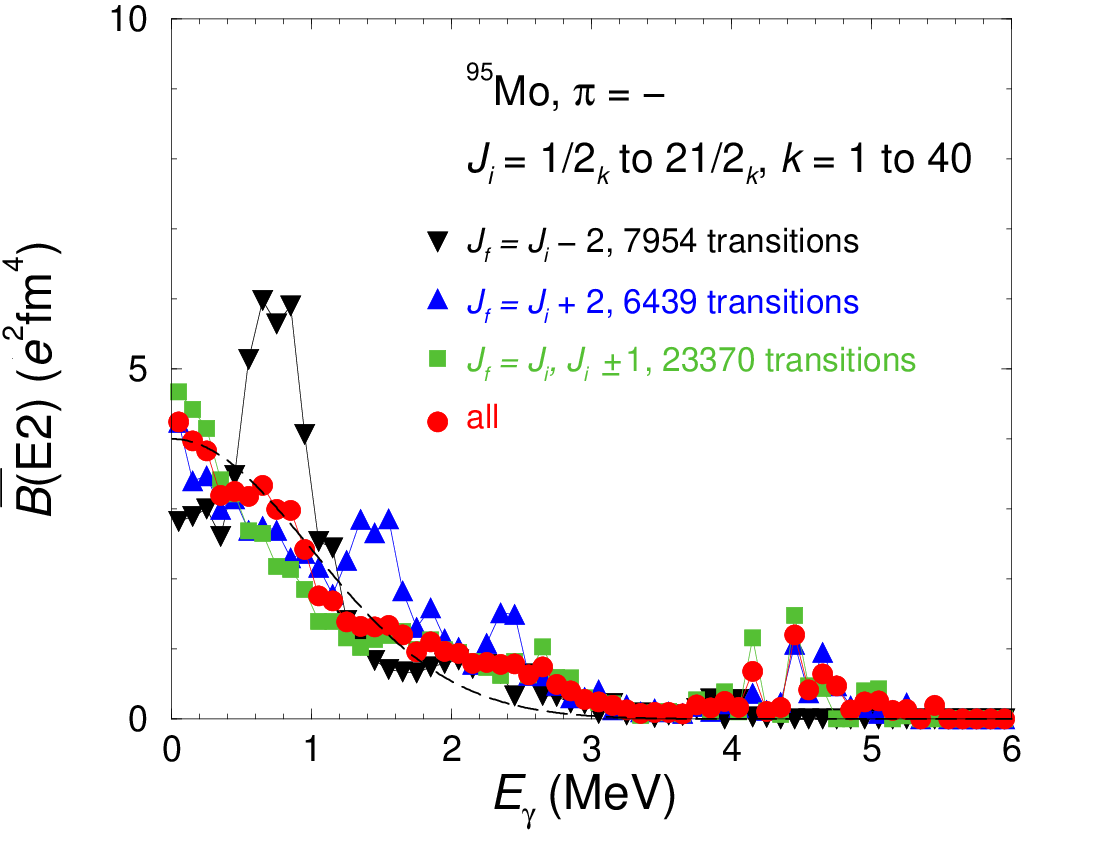,width=8cm}
\caption{\label{fig:95MoNEgE2}(Color online) As Fig.~\ref{fig:95MoPEgE2}, but
for negative-parity states.} 
\end{figure}

The analogous plots for positive-parity states and negative-parity states in
$^{95}$Mo are shown in Figs.~\ref{fig:95MoPEgE2} and
\ref{fig:95MoNEgE2}, respectively. The distributions in this odd-mass $N$ = 53
nuclide look similar to the ones in the even-mass $N$ = 52 neighbor and are 
created by states analogous to the ones in $^{94}$Mo. The low-energy parts of
the distributions of all transitions may be approximated by Gau{\ss} curves
with parameters of $B_0$ = 5.0 $e^2$fm$^4$, $\sigma$ = 1.0 MeV for positive
parity and $B_0$ = 4.0 $e^2$fm$^4$, $\sigma$ = 1.0 MeV for negative parity,
which are very close to the corresponding values in $^{94}$Mo.

\section{$E2$ strength functions}
\label{sec:E2sf}

$E2$ strength functions have been deduced from the $\overline{B}(E2)$
distributions including all transitions in a way analogous to the one described
in Ref.~\cite{sch13}. To calculate the $E2$ strength function the relation
$f_2(E_\gamma) = 0.80632 \times 10^{-12}$ $\overline{B}(E2,E_\gamma)$ 
$\rho(E_i)$ was used, where $\rho(E_i)$ is the level density in MeV$^{-1}$
at the energy of the initial state. The $f_2(E_\gamma)$ values were deduced in
energy bins as done for the $\overline{B}(E2)$ values. The level densities
$\rho(E_i,\pi)$ were determined by counting the calculated levels
within energy intervals of 1 MeV for the two parities separately. The total
level densities $\rho(E_i)$ are well reproduced by the constant-temperature
expression $\rho(E_i) = \rho_0 \exp{(E_i/T_\rho)}$ for $E_i <$ 5 MeV. For
higher energies the level density decreases with excitation energy, which is
due to missing levels at high energy in the present configuration space and
spin range. The parameters of the expression for $\rho$ are 
$\rho_0$ = 1.37 MeV$^{-1}$, $T_\rho$ = 0.67 MeV for $^{94}$Mo and
$\rho_0$ = 1.90 MeV$^{-1}$, $T_\rho$ = 0.54 for $^{95}$Mo \cite{sch13}. 

\begin{figure}
\epsfig{file=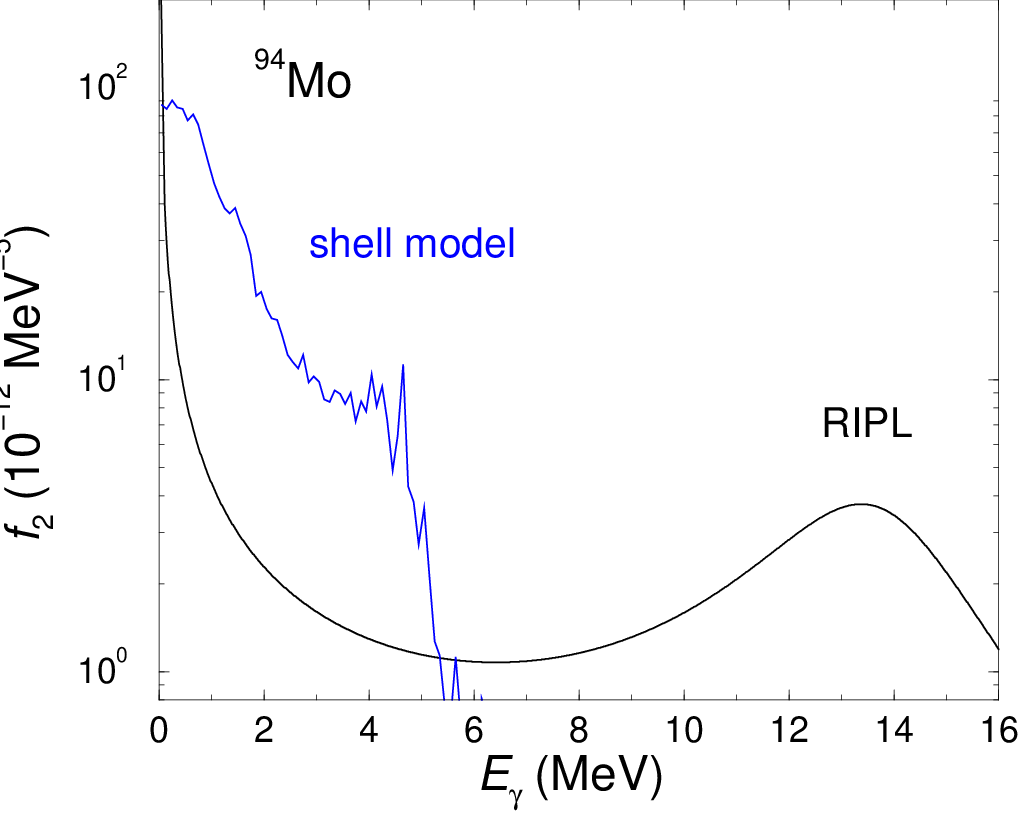,width=8cm}
\caption{\label{fig:94MoEgf2}(Color online) $E2$ strength function for
$^{94}$Mo deduced from the present shell model calculations (blue line) and 
the $E2$ strength function according to the expression given in the RIPL
handbook (black curve) \cite{bel06}.}. 
\end{figure}

\begin{figure}
\epsfig{file=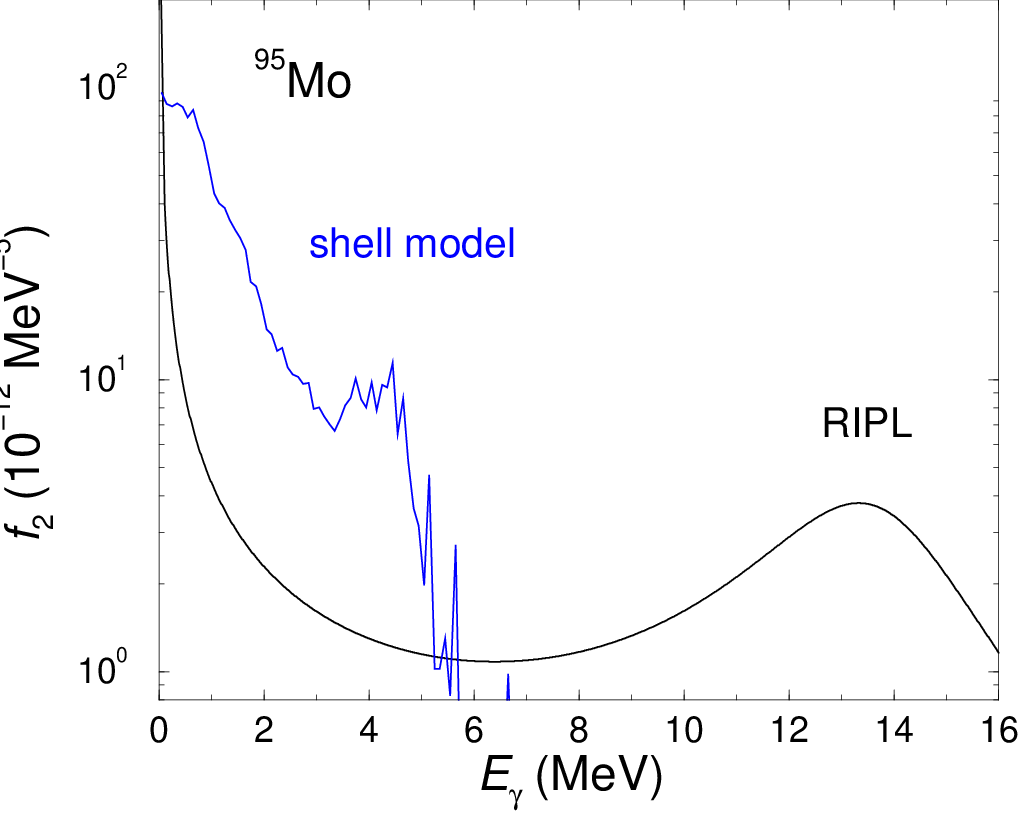,width=8cm}
\caption{\label{fig:95MoEgf2}(Color online) As Fig.~\ref{fig:94MoEgf2}, but
for $^{95}$Mo.}
\end{figure}

The total $E2$ strength functions for $^{94}$Mo and $^{95}$Mo are shown in 
Figs.~\ref{fig:94MoEgf2} and \ref{fig:95MoEgf2}, respectively. As seen for the
$\overline{B}(E2)$ distributions, the $E2$ strength functions are bell-shaped
at low energy below about 2 MeV. This is different from the low-energy energy
behavior of the $M1$ strength functions calculated within the shell model which
steadily increase toward $E_\gamma$ = 0 \cite{sch13}.

For comparison, the curves calculated according to the phenomenological
expression recommended in the RIPL handbook \cite{bel06} for the $E2$ strength
function are plotted in Figs.~\ref{fig:94MoEgf2} and \ref{fig:95MoEgf2}. At
$E_\gamma$ = 0 these curves have an unphysical pole in contrast to the finite
maximum resulting from the present shell-model calculations. At medium energies
the low-energy tails of the Lorentz curves underestimate the $E2$ strength
predicted in the shell-model calculations by more than one order of magnitude.

\section{Summary}
\label{sec:conclu}

A large number of $E2$ transitions between excited states up to $J$ = 10 in
$^{94}$Mo and $^{95}$Mo has been calculated using the shell model. At low
transition energy below about 2 MeV, the distributions of average $B(E2)$
values are dominated by the large number of transitions between states with
$J_f = J_i, J_i \pm 1$. These are the transitions with large average $B(M1)$
values discussed in Ref.~\cite{sch13}. The corresponding states contain large
components of the configuration $\pi(0g_{9/2}^2) \nu(1d_{5/2}^2)$. The strength
functions deduced from the average $E2$ strengths increase toward zero
transition energy and show a finite maximum of a Gau{\ss}-like shape. This is
in contrast to the pole of the phenomenological expression recommended in the
reaction data base RIPL. In the medium-energy range up to about 6 MeV the
average $E2$ strength predicted by the shell-model calculations shows a
complicated structure and is by orders of magnitude greater than the low-energy
tail of the phenomenological expression. This part may miss components caused
by collective excitations and may show a different behavior in nuclides that
are more distant from shell closures than the ones studied in this work. The
continuation of the strength to higher energy beyond about 6 MeV remains an
open question. The possible influence of the low-energy shape of the $E2$
strength functions on reaction rates may be tested by implementing these
strength functions in statistical reaction codes.

\section{Acknowledgments}

Helpful discussions with B. A. Brown and S. Frauendorf are gratefully
acknowledged.

\end{document}